\documentclass[final,5p,times,twocolumn]{elsarticle}

\usepackage{graphicx,amsmath,amssymb,bm,multirow}
\usepackage{amsthm}
\usepackage{amsfonts}

\usepackage[usenames,dvipsnames]{color}
\usepackage[colorlinks,linkcolor=blue,anchorcolor=blue,citecolor=blue]{hyperref}

\newcommand{\br}{{\mathbf{r}}}
\newcommand{\elemA}[2]{\ensuremath{{}^{#1}}\textrm{#2}}

\usepackage[nodots]{numcompress}

\begin{document}

\begin{frontmatter}

\title{Does a proton ``bubble" structure exist in the low-lying states of $^{34}$Si?}

\author{J. M. Yao\fnref{1,2}}
\author{H. Mei\fnref{1}}
\author{Z. P. Li\fnref{1}}
\address[1]{School of Physical Science and Technology,
             Southwest University, Chongqing 400715, China}
\address[2]{Physique Nucl\'eaire Th\'eorique, Universit\'e Libre de Bruxelles, C.P. 229, B-1050 Bruxelles, Belgium}
%
\begin{abstract}
The possible existence of a ``bubble" structure in the proton density of $^{34}$Si
has recently attracted a lot of research interest. To examine the existence of the ``bubble" structure in low-lying states, we establish a relativistic version of configuration mixing of both particle number and angular momentum projected quadrupole deformed mean-field states and apply this state-of-the-art beyond relativistic mean-field method to study the density distribution of the low-lying states in $^{34}$Si. An excellent agreement with the data of low-spin spectrum and electric multipole transition strengths is achieved without introducing any parameters. We find that the central depression in the proton density is quenched by dynamic quadrupole shape fluctuation, but not as significantly as what has been found in a beyond non-relativistic mean-field study. Our results suggest that the existence of proton ``bubble" structure in the low-lying excited $0^+_2$ and $2^+_1$ states is very unlikely.

\end{abstract}


\end{frontmatter}

%
%

{\em Introduction}.$-$ In recent years, there has been a growing interest in searching for ``bubble" nuclei, in which the density in the center vanishes or is significantly lower than saturation density. The ``bubble" nuclei are of particular interest because they have different mean-field potentials from those of normal nuclei with Fermi-type density distribution. In particular, the splitting of spin-orbit partners located mainly at the nuclear center is reduced in the ``bubble" nuclei. Some modern parametrizations of the relativistic mean field (RMF) \cite{Todd-Rutel2004,Chu2010} and of the Skyrme energy density functional (EDF) \cite{Khan2008,Grasso2009,Wang11}, as well as the mean-field calculation using semi-realistic interactions~\cite{Nakada12} predict a hollow proton density for $\elemA{34}{Si}$ and some neutron-rich Ar isotopes. At the time being, $\elemA{34}{Si}$ stands out as the only candidate on which many different studies agree. The possible proton ``bubble'' structure of this nucleus has also been suggested as an explanation for the recently observed reduction of the spin-orbit splitting using the transfer reactions $\elemA{36}{S}(d,p)\elemA{37}{S}$ and $\elemA{34}{Si}(d,p)\elemA{35}{Si}$ \cite{Burgunder2011}. Recently, the intruder $0^+_2$ state and its weak electric monopole transition to the ground-state were measured~\cite{Rotaru12}. The results support the coexistence structure of a spherical ground-state with a large deformed $0^+_2$ state in $^{34}$Si. The spectroscopy of low-lying states provides a strong test of the nuclear structure models that have been used to study the density profiles.

Most recently, the stability of ``bubble" structure against dynamical effects in $^{34}$Si has been examined in the framework of a particle-number (PN) and angular-momentum (AM) projected generator coordinate method (GCM) based on Hartree-Fock-Bogoliubov (HFB) states with axial quadrupole deformation using the non-relativistic Skyrme force SLy4~\cite{Yao12}. It has been shown that the dynamic effect of quadrupole shape fluctuation significantly altered the radial density profile, and brought it closer to a Fermi-type density distribution. We noted, however, that the spectroscopic properties of the observed low-lying states were not reproduced very well~\cite{Yao12}. The calculated two lowest $0^+$ states were too strongly mixed, which might overestimate the effect of shape mixing on the density of ground-state. Moreover, in the context of projected GCM based on self-consistent mean-field approaches, the density profiles of excited states have not been studied yet. The existence of the proton ``bubble" structure in the low-lying excited states of $^{34}$Si is not known.

During the past decades, the RMF theory has achieved great success in describing many nuclear phenomena for both stable and exotic nuclei over the entire nuclear chart with a few universal parameters~\cite{Reinhard89,Ring96,Vretenar05,Meng06}. It incorporates many important relativistic effects, such as the presence of large Lorentz scalar and vector fields with approximately equal magnitude and opposite sign. This leads to a new saturation mechanism via the difference between the scalar and vector densities, and naturally to the large spin-orbit interaction needed for the understanding of magic numbers in finite nuclei. The aim of the present Letter is to address the above issue using a beyond RMF approach.

{\em The method}.$-$ To this end, we extend the beyond RMF method presented in Ref.~\cite{Yao10} by including additionally PN projection and restricting it to axially deformed states by imposing triaxial deformation parameter $\gamma$ to be $0^\circ$ or $180^\circ$. In this case, the wave function of nuclear low-lying state is given by the superposition of both PN and AM projected RMF wave functions constrained to have different intrinsic axial deformations $\beta$,
 \begin{equation}
 \vert \Psi^{JNZ}_\alpha\rangle
 =\sum_\beta f^{JNZ}_\alpha(\beta)\hat P^J_{M0} \hat P^N\hat P^Z\vert \Phi(\beta)\rangle,
 \end{equation}
 with $\hat P^J_{M0}$, $\hat P^N$, $\hat P^Z$ being the projection operators onto good number of  angular momentum, neutrons and protons, respectively. The weight factors $f_\alpha^{JNZ}(\beta)$ and the energies $E_\alpha^{J}$ of the states $\vert \Psi^{JNZ}_\alpha\rangle$ are the solutions of the Hill-Wheeler-Griffin equation~\cite{Ring80}
\begin{equation}
\label{eq_GCM:20}
\sum_{\beta'} \left[ \mathcal{H}^{J}(\beta,\beta') - E_\alpha^{J}\mathcal{N}^{J}(\beta,\beta')
          \right] \, f_\alpha^{JNZ}(\beta')
= 0,
\end{equation}
where $\mathcal{N}^{J}(\beta,\beta')=\langle \Phi(\beta)\vert\hat P^J_{00} \hat P^N\hat P^Z\vert \Phi(\beta')\rangle$ and $\mathcal{H}^{J}(\beta,\beta')=\langle \Phi(\beta)\vert\hat H\hat P^J_{00} \hat P^N\hat P^Z\vert \Phi(\beta')\rangle$ are the norm kernel and the energy kernel, respectively. In the calculations, the energy overlap in the energy kernel is taken with the same functional form as the nuclear mean-field energy, but replacing the densities and currents with mixed ones~\cite{Yao10}. We note that in recent years several similar beyond mean-field methods (GCM+PNP+AMP) are developed by different groups based on either non-relativistic or relativistic EDFs~\cite{Niksic06,Bender2008,Rodriguez10}.

Once the wavefunctions of low-lying states are found, it is straightforward to calculate the density distribution in $r$-space. The density distribution for the state $\vert \Psi^{JNZ}_\alpha\rangle$ is found as~\cite{Yao13}
\begin{eqnarray}
\label{eq_GCM:70}
\rho^{J\alpha}(\br)
  & = &   \sum_{\beta\beta'}  f^{JZN}_{\alpha} (\beta')f^{JZN}_\alpha(\beta)
  \sum_{\lambda}  (-1)^{2\lambda}Y_{\lambda0}(\hat \br) \nonumber\\
 &&\times
 \langle  J0,\lambda 0\vert J 0\rangle  \sum_{K_2}(-1)^{K_2}\langle JK_2,\lambda -K_2\vert J0\rangle \nonumber \\
        &   & \times
  \int d\hat\br^\prime \rho^{JK_2}_{\beta'\beta}(\br^\prime)Y^\ast_{\lambda K_2}(\hat \br^\prime),
\end{eqnarray}
where the $\rho^{JK_2}_{\beta'\beta}(\br)$ is defined as
\begin{eqnarray}
  \label{rho_JKNZ}
  \rho^{JK_2}_{\beta'\beta}(\br)
 &\equiv& \dfrac{2J+1}{2} \int^\pi_0 d\theta\sin(\theta) d^{J\ast}_{K_20}(\theta)\nonumber\\
 &&\times
 \langle  \Phi(\beta^\prime) \vert \sum_{i}\delta(\br-\br_i) e^{i\theta\hat J_y} \hat P^{N}\hat P^{Z}  \vert \Phi(\beta)\rangle.
\end{eqnarray}
 The sum over index $i$ represents summation over all occupied single-particle states for neutrons or protons.

 The reduced electric quadrupole ($E2$) transition strengths between low-lying states are given by
\begin{eqnarray}
\lefteqn{
B(E2;J_i,\alpha_i\rightarrow J_f,\alpha_f)
} \nonumber\\
& = & \dfrac{1}{2J_i+1} \, \Big|
      \sum_{\beta^\prime,\beta} f^{J_fNZ\ast}_{\alpha_f}(\beta^\prime) \,
      \langle J_f \beta^\prime || \hat Q_{2} || J_i \beta\rangle \,
      f^{J_iNZ}_{\alpha_i}(\beta) \Big|^2
\end{eqnarray}
and are calculated directly in the laboratory frame without approximation.
The reduced matrix elements in the above expression are determined as
\begin{eqnarray}
\lefteqn{
\langle J_f \beta^\prime|| \hat Q_{2}|| J_i \beta\rangle
}
\nonumber\\
& = &
  \dfrac{(2J_f+1)(2J_i+1)}{2}
 \sum_{M=-2}^{+2}
 \begin{pmatrix}
 J_f & 2 &  J_i \\
  0  & M &  -M
 \end{pmatrix} \nonumber\\
 &&\times \int^\pi_0 \! d\theta \, \sin(\theta) \, d^{J_i\ast}_{-M0}(\theta)
 \langle  \Phi(\beta^\prime)| \hat Q_{2M} e^{i\theta\hat J_y} \hat P^N\hat P^Z  |  \Phi(\beta) \rangle \, , \nonumber\\
\end{eqnarray}
where $\hat Q_{2M} \equiv e \, r^2 \, Y_{2M}$ is the electric quadrupole
moment operator.  The nuclear matrix element entering the
electric monopole decay from $| \Psi^{JNZ}_\alpha\rangle$  to $| \Psi^{JNZ}_{\alpha^\prime} \rangle$ through the emission of conversion electrons is determined by
\begin{equation}
\label{rho:E0}
\rho^2(E0; J_\alpha \to J_{\alpha^\prime})
= \left| \frac{\langle \Psi^{JNZ}_{\alpha^\prime} | e \sum_p r^2_p | \Psi^{JNZ}_\alpha \rangle}{eR^2}
  \right|^2 \, ,
\end{equation}
where $p$ is an index running over all proton single-particle states.
The radius $R$ is given by $R = 1.2 A^{1/3}$~fm.
Since the electric transition matrix elements are calculated in the
full model space of occupied single-particle states, there is no need to
introduce effective charges, and the  bare charge for protons is used instead.

{\em Numerical details}.$-$ In the RMF calculations, parity, $x$-simplex symmetry, and time-reversal invariance are imposed. The Dirac equation is solved by expanding in the basis of eigenfunctions of a three-dimensional harmonic oscillator in Cartesian coordinate with 10 major shells, which are found to be sufficient to obtain reasonably converged results for $^{34}$Si. A constraint on the axial mass quadrupole moment $\langle Q_{20} \rangle = \sqrt{\frac{5}{16\pi}}\langle 2 z^2 - x^2 - y^2 \rangle$ is used to generate mean-field states $| \Phi(\beta) \rangle$ with different
intrinsic deformations $\beta= \frac{4\pi}{3AR^2} \langle Q_{20} \rangle$. The point-coupling type of relativistic effective force PC-PK1~\cite{Zhao10} is adopted. Pairing correlations between nucleons are treated with the BCS approximation using a density-independent $\delta$ force with a smooth cutoff factor~\cite{Krieger90}. The strength parameters of the pairing force are $V_n=-349.5$ and $V_p=-330$ MeV$\cdot$fm$^3$ for neutrons and protons respectively.

The Gauss-Legendre quadrature is used for integrals over the Euler angle $\theta$ and gauge angle $\varphi_{\tau=n,p}$ in the calculations of the AM+PN projected norm and Hamiltonian kernels. The number of mesh points in the interval $[0, \pi]$ for $\theta$ and $\varphi_\tau$  are chosen as 14 and 9 respectively. The pfaffian method~\cite{Robledo09} is implemented to calculate the norm overlap, the phase of which can be uniquely determined in this way.

\begin{figure}[]
\centering
\includegraphics[width=8.5cm]{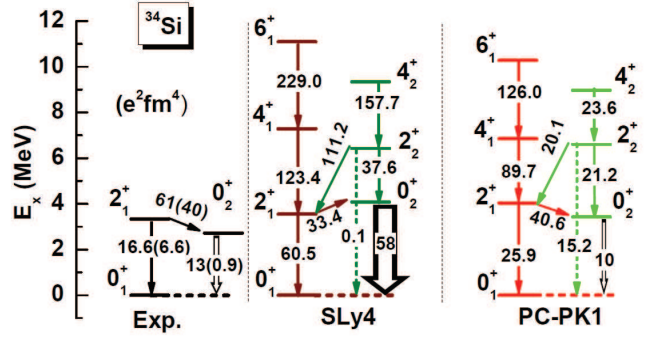}
\caption{(Color online.) Comparison between the (left) experimental data~\cite{Rotaru12},
(middle) the predictions by the SLy4 force~\cite{Yao12}, and
(right) the ones by the PC-PK1 force for the low-lying states in $^{34}$Si.
The electric monopole ($E0$) transition between the first two $0^+$ states
is indicated with the value of $\rho^2(E0; 0^+_2\to 0^+_1)\times 10^3$. The electric quadrupole ($E2$) transition is indicated with the value in units of $e^{2}$ fm$^{4}$. }
\label{spectrum}
\end{figure}

{\em Results and discussions}.$-$ Figure~\ref{spectrum} displays the low-lying states in $^{34}$Si from the beyond mean-field calculations using both the non-relativistic SLy4 force and the relativistic PC-PK1 force, in comparison with the recently measured data~\cite{Rotaru12}. As we have discussed in Ref.~\cite{Yao12} that the calculations using the SLy4 force reproduced the energy of the recently observed low-energy $0^+_2$ state and the interband $B(E2; 2^+_1 \rightarrow 0^+_2)$ value rather well. However, the electric monopole transition strength $\rho^2(E0; 0^+_2\rightarrow0^+_1)$ and the in-band $B(E2; 2^+_1 \rightarrow 0^+_1)$ value were overestimated, which indicates that the two lowest $0^+$ GCM states are too strongly mixed in the calculation. In the relativistic calculations using the PC-PK1 force, however, both the excitation energies and $E0$, $E2$ transition strengths of the low-lying states are reproduced quite well, which gives us a strong motivation to study the density distributions of low-lying states using the wave functions obtained in the relativistic calculations.

\begin{figure}[]
\centering
\includegraphics[width=8cm]{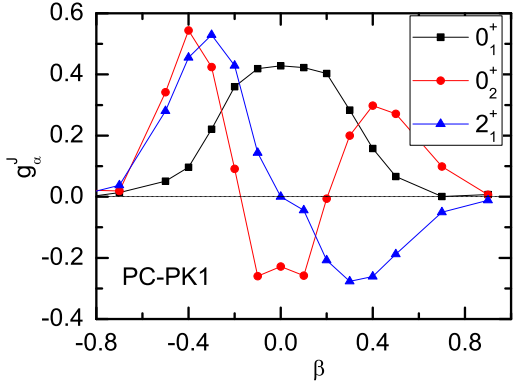}
\caption{(Color online.) Collective wave functions [cf.Eq.(\ref{eq_GCM:30})] of the $0^+_1, 0^+_2, 2^+_1$ states in $^{34}$Si.}
\label{gcmwfs}
\end{figure}

Figure~\ref{gcmwfs} displays the distribution of the collective wave functions $g_\alpha^{J}(\beta)$ in deformation $\beta$ degree of freedom for the $0^+_1, 0^+_2, 2^+_1$ states, among which the electric multipole transitions have been observed~\cite{Rotaru12}, where the $g_\alpha^J(\beta)$ are related to the weight function $f^{JNZ}_\alpha$ by the following relation,
\begin{equation}
\label{eq_GCM:30}
g_\alpha^{J}(\beta)
= \sum_{\beta'} \big( \mathcal{N}^{J}(\beta,\beta') \big)^{1/2} f^{JNZ}_\alpha(\beta'),
\end{equation}
and are orthonormal to each other. The shapes of collective wave functions for the $0^+_1$ and $0^+_2$ states are similar to those obtained in Ref.~\cite{Yao12} from the calculations using the SLy4 force. However, there are differences in detail. The wave function of ground-state is more concentrated around the spherical shape in the relativistic case, while that of $0^+_2$ state has more weight in large deformed configurations (dominated by the intruder neutron $f_{7/2}$ orbital). As a consequence, compared with the SLy4 results, the $\rho^2(E0;0^+_2\rightarrow0^+_1)$ and $B(E2;2^+_1 \rightarrow 0^+_1)$ values are reduced and become closer to the data in the beyond RMF calculations.

\begin{figure}[]
\centering
\includegraphics[width=8cm]{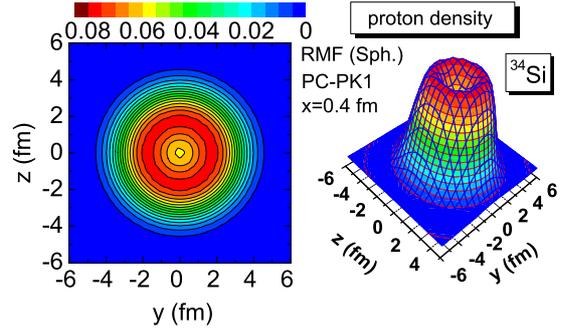}
\caption{(Color online.) Density distribution of mean-field spherical state in $y$-$z$ plane at $x=0.4$ fm from the RMF calculation using the PC-PK1 force for $^{34}$Si (in fm$^{-3}$).}
\label{density}
\end{figure}

\begin{figure}[]
\centering
\includegraphics[width=7cm]{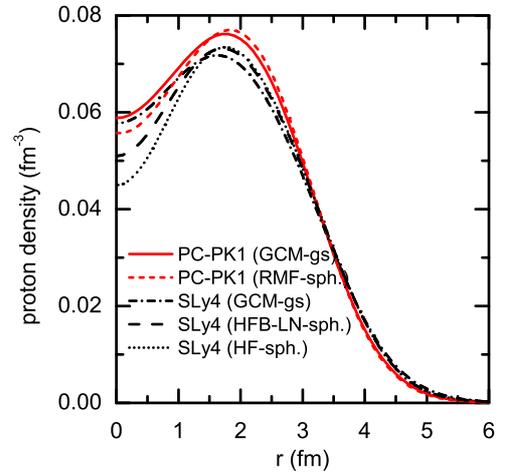}
\caption{(Color online.) Comparison of proton density distributions of ground-state from both mean-field and beyond mean-field calculations for $^{34}$Si. The non-relativistic results, taken from Ref.~\cite{Yao12}, are given for comparison.}
\label{compdens}
\end{figure}

The density distribution of protons in $^{34}$Si from the RMF calculation is displayed in Fig.~\ref{density}, which shows an evident central depletion. This density distribution corresponds to the spherical mean-field configuration, for which case, the occupation probability of proton $2s_{1/2}$ orbital is zero due to pairing collapse. The mixing of configurations of different intrinsic shapes resulting from dynamic fluctuation in quadrupole deformations can alter the occupancy of the $2s_{1/2}$ orbital and therefore change the density distribution in the center. Figure~\ref{compdens} displays the comparison of mean-field and beyond mean-field (GCM+PNP+AMP) calculated density distributions for the ground-state in $^{34}$Si. It shows that the quenching of the depletion of proton density in the center by shape mixing is a common feature in the beyond mean-field calculations. However, the beyond RMF calculation gives a smaller quenching effect than that the one found in the beyond non-relativistic mean-field calculation, which can be mainly attributed to the difference in the distribution of wave function in the collective space, i.e., the wave function of ground-state in this work is more concentrated around the spherical shape than that given in Ref.~\cite{Yao12}.

\begin{figure}[]
\centering
\includegraphics[width=7cm]{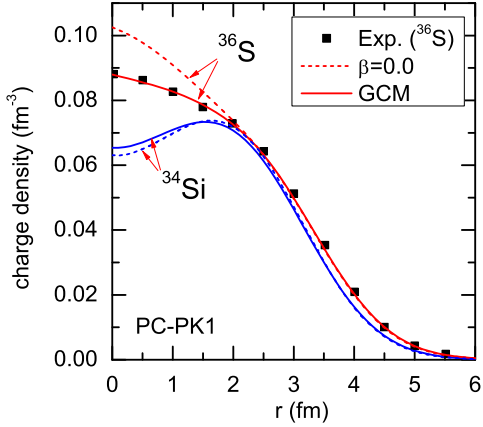}\vspace{-0.25cm}
\caption{(Color online.) Charge density distributions in $^{34}$Si and $^{36}$S from the relativistic calculations using the PC-PK1 force. The experimental data for $^{36}$S are taken from Ref.~\cite{Rychel83}}.
\label{charge-density}
\end{figure}

Figure~\ref{charge-density} displays the comparison of charge density distributions in $^{34}$Si and $^{36}$S from the relativistic calculations using the PC-PK1 force. The charge density is calculated by convolution of the proton density with a Gaussian form factor with a proton size $a = \sqrt{2/3} \langle r^2 \rangle^{1/2}_p = 0.65$~fm, which for spherically symmetric density distributions leads to~\cite{Negele70}
\begin{equation}
\label{charge-sph}
\rho_{\rm ch} (r)
  =
    \frac{1}{a\sqrt{\pi}}
    \int \! dr' r' \, \rho_p(r') \,
    \left[
      \frac{e^{-(r-r')^2/a^2}}{r} - \frac{e^{-(r+r')^2/a^2}}{r}
    \right] \, .
\end{equation}
The charge density of mean-field result ($\beta=0.0$) for $^{36}$S is much higher than the experimental data. This phenomenon is also observed in other RMF calculations using the NL3 and DD-ME2 forces~\cite{Grasso2009}. Figure~\ref{charge-density} demonstrates that after including the dynamic correlation effects in the projected GCM calculation, the charge density of $^{36}$S is in excellent agreement with the data, which gives us confidence in the prediction for the charge density of $^{34}$Si.

\begin{figure}[]
\centering
\includegraphics[width=8cm]{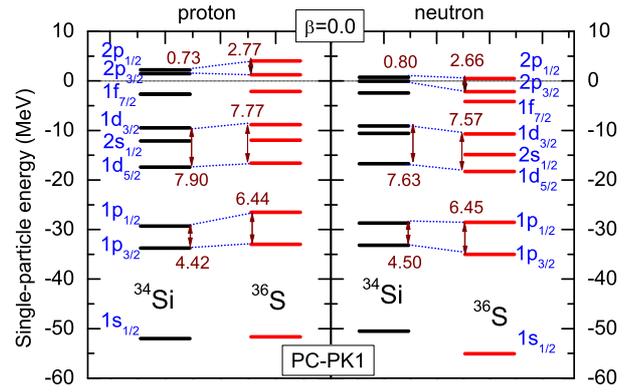}\vspace{-0.5cm}
\caption{(Color online.) Single-particle energy spectra for protons and neutrons corresponding to the spherical configuration of $^{34}$Si and $^{36}$S from the RMF calculations using the PC-PK1 force. The size of spin-orbit splitting is indicated with the value in units of MeV. }
\label{spe}
\end{figure}

In the mean-field approaches, a central depletion of the proton density in $^{34}$Si could induce a non-zero density derivative in the center and thus reduce the strength of the spin-orbit interaction for the inner orbits. This has been suggested as an explanation for the reduction of neutron $2p_{3/2}-2p_{1/2}$ splitting between $^{37}$S and $^{35}$Si inferred from transfer
reactions~\cite{Burgunder2011}. To examine this effect at the mean-field level, we plot in Fig.~\ref{spe} the single-particle energy spectra for protons and neutrons corresponding to the spherical configuration of $^{34}$Si and $^{36}$S from the RMF calculations using the PC-PK1 force.
As expected, the splitting of spin-orbit doublets $2p_{3/2}-2p_{1/2}$ is reduced significantly from 2.77(2.66) MeV to 0.73(0.80) MeV for protons (neutrons) when going from $^{36}$S to $^{34}$Si. It should be noted that the spin-orbit interaction is emerging naturally from the derivative of vector and scalar fields in the RMF approaches and no adjustable parameter is introduced. However, one should not compare these values directly with those inferred from transfer reactions, as discussed in Ref.~\cite{Yao12}. Moreover, as shown in Fig.~\ref{charge-density} that the pure RMF calculation overestimates the central density in $^{36}$S, and therefore enhances the spin-orbital splitting. Furthermore, the neutron $2p_{3/2}$ level in $^{34}$Si is weakly bound at -0.063 MeV, whereas the neutron $2p_{1/2}$ level in $^{34}$Si and $^{36}$S is unbound at +0.74 MeV and +0.48 MeV respectively, in which case, the coupling to the continuum has to be carefully taken into account, which is beyond the scope of present study.

\begin{figure}[]
\centering
\includegraphics[width=7cm]{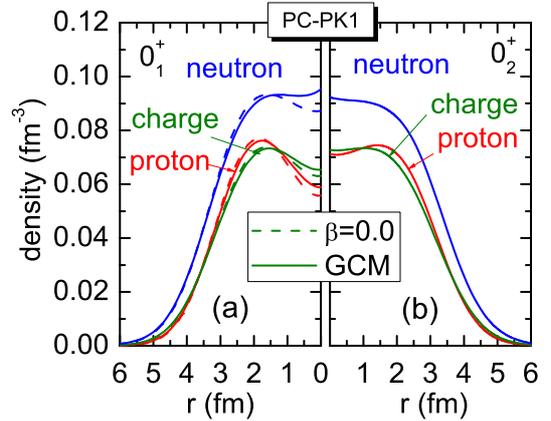}
\caption{(Color online.) Density distributions of neutrons, protons and charges for the first two $0^+$ states (solid lines), in comparison with the mean-field calculated results for the spherical configuration (dashed lines).}
\label{gcmdens}
\end{figure}

Figure~\ref{gcmdens} displays the density distributions of neutrons, protons and charges for the first two $0^+$ states, in comparison with the mean-field calculated results of pure spherical configuration. It shows clearly that the densities of both neutrons and protons in the ground-state are closer to the Fermi-type distribution in the projected GCM calculation, in comparison with the pure RMF calculation ($\beta=0.0$). In particular, for the $0^+_2$ state, the central depression in the density distribution is not visible.

\begin{figure}[]
\centering
\includegraphics[width=8cm]{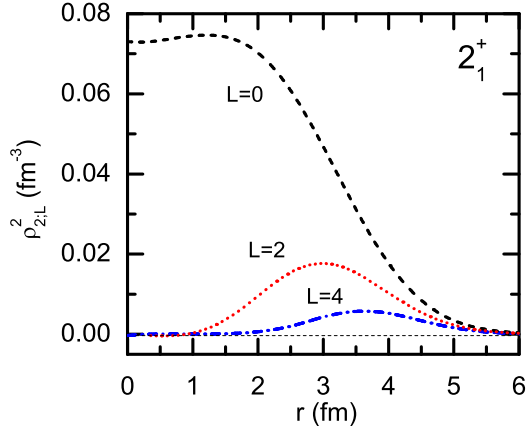}
\caption{(Color online.) Radial transition densities ($L=0, 2, 4$) of the $2^+_1$ state in $^{34}$Si. The $L=0$ component is scaled by dividing a factor of $\sqrt{4\pi}$.}
\label{densj2}
\end{figure}

For non-zero spin states, the density is deformed, in which case, it is convenient to define a radial (reduced) transition density with angular momentum transfer $L$~\cite{Yao13},
\begin{eqnarray}
 \label{trans-dens-JJ}
 \rho^{J}_{J;L}(r)
 &=&\langle J0 L0\vert J0\rangle^{-1}
 \int d\hat\br  \rho^{J\alpha}(\br)  Y_{L0}(\hat\br),
\end{eqnarray}
where $\langle J0 L0\vert J0\rangle$ is a Clebsch-Gordan coefficient. We note that the integration of the radial transition density $\rho^{J}_{J;0}(r)$, multiplied with $r^{4}$ over the radial coordinate $r$ is related to the mean-square radius of the state $\langle r^2\rangle_{J\alpha}$ by the relation $\int \rho^{J}_{J;0}(r) r^4 dr=\langle r^2\rangle_{J\alpha}/\sqrt{4\pi}$. We find that the mean-square radius of the first three low-lying states are respectively $\langle r^2\rangle_{0^+_1}=3.09$ fm, $\langle r^2\rangle_{0^+_2}=3.15$ fm, and $\langle r^2\rangle_{2^+_1}=3.21$ fm. Moreover, the spectroscopic quadrupole moment of $2^+_1$ state is 9.48 $e$ fm$^2$, which indicates that the $2^+_1$ state is predominately oblate deformed.

Figure~\ref{densj2} displays the radial transition densities ($L=0, 2, 4$) of the $2^+_1$ state in $^{34}$Si. It is clearly shown that the central depression does not exist in the $L=0$ component of radial transition density. The disappearance of ``bubble" structure in the $0^+_2$ and $2^+_1$ states is due to the mixing of (prolate) deformed configurations, in which, the component originated from the proton spherical $2s_{1/2}$ orbital has a sizable occupation probability.

\begin{table}[]
\tabcolsep=4pt
\caption{
\label{factor} The central $(\rho_{\text{cent}})$ and maximal $(\rho_{\text{max}})$ proton densities, as well as the depletion factors $F_{\text{max}}$ and $F_{\text{sat}}$ for the proton and charge densities in $^{34}$Si from the calculations using the PC-PK1 force.}
\begin{center}
\begin{tabular}{l|cccc|cc}
\hline \noalign{\smallskip}
  state   &
$\rho_{\text{cent}}$ &
$\rho_{\text{max}}$  &
$F_{\text{max}}$     &
$F_{\text{sat}}$    & $F_{\text{max}}(\rm ch.)$ &  $F_{\text{sat}}(\rm ch.)$ \\
\noalign{\smallskip} \hline \noalign{\smallskip}
 $\beta=0.0$     & 0.056 & 0.077  & 0.27  &0.15 & 0.14  & 0.04 \\
 $0^+_1$         & 0.059 & 0.076  & 0.22  &0.11 & 0.11  & 0.01 \\
 $0^+_2$         & 0.071 & 0.074  & 0.04  &$-$0.08 & 0.01  & $-$0.10 \\
 $2^+_1$ ($L=0$) & 0.073 & 0.075  & 0.03  &$-$0.11 &     &    \\
\noalign{\smallskip} \hline
\end{tabular}
\end{center}
\end{table}

Table~\ref{factor} lists the values of the central and maximal proton densities, and the depletion factors $F_{\textrm{max}}$ and $F_{\textrm{sat}}$~\cite{Yao12} for the proton and charge (ch.) densities in $^{34}$Si from the calculations using the PC-PK1 force, where
\begin{equation}
\label{eq:F}
F_{\text{max}}
\equiv \frac{\rho_{\textrm{max}}-\rho_{\textrm{cent}}}
            {\rho_{\textrm{max}}},~~
F_{\text{sat}}
\equiv \frac{ \rho_{\textrm{sat}} - \rho_{\textrm{cent}}}
            {\rho_{\textrm{sat}}},
\end{equation}
with $\rho_{\textrm{sat}}$ being the saturation value of the proton density, $\rho_{\text{sat}} = (14/34) \times0.16$~fm$^{-3} = 0.066$~fm$^{-3}$ for $\elemA{34}{Si}$. The beyond mean-field effects, dominated by the shape mixing, reduce the depletion factor $F_{\text{max}}$ from 0.14 to 0.11 and $F_{\text{sat}}$ from 0.04 to 0.01 for the charge density, which are compared with the results of  the SLy4 force, i.e., $F_{\text{max}}$ from 0.16 to 0.09 and $F_{\text{sat}}$ from 0.09 to 0.04. For the observed low-lying $0^+_2$ and $2^+_1$ states, the depletion factor is close to zero, which demonstrates the disappearance of proton ``bubble" structure quantitatively. Finally, we point out that the implementation of Lipkin-Nogami (LN) prescription for pairing correlation in our RMF calculations would bump protons onto $2s_{1/2}$ orbital and consequently reduce the ``bubble" structure further.

{\em Summary and conclusions}.$-$ We have established a beyond RMF method by mixing of both particle number and angular momentum projected quadrupole deformed states and applied it to study the excitation energies, electric multipole transition strengths and the density distributions for the low-lying states in $^{34}$Si. The recently observed spectroscopic data have been reproduced quite well without introducing any parameters, which provides us confidence on the reliability of the wave functions and the resultant density distributions. Moreover, we have found that for $^{36}$S, the dynamic correlation effects have significant influence on the central charge density. After considering these effects in the beyond RMF calculation, the charge density distribution is in excellent agreement with the data, which gives us more confidence in the prediction for the charge density of $^{34}$Si. We have found that the central depression in the proton and charge densities of $^{34}$Si is quenched by dynamic quadrupole shape fluctuation, but not as significantly as what has been found in a non-relativistic study using the SLy4 force. Our studies suggest that the existence of a proton ``bubble" structure in the low-lying excited $0^+_2$ and $2^+_1$ states is very unlikely. These findings are hoped to be examined in the new generation of electron-RIB colliders SCRIT (Self Confining Radioactive Isotope Target in Japan~\cite{Suda09}) and ELISe (ELectron-Ion Scattering in a storage ring in Germany~\cite{Antonov11}) in the near future.

%
\section*{Acknowledgments}

J.M.Y. acknowledges S. Baroni, M. Bender, K. Hagino and P.-H. Heenen for their stimulating discussions and useful comments on this manuscript. This work was partly supported by the NSFC under Grant Nos. 10947013, 11105110 and 11105111, the European Union's Seventh Framework Programme ENSAR under grant agreement n262010, the Fundamental Research Funds for the Central Universities (XDJK2010B007 and XDJK2011B002), and the NSF of Chongqing cstc2011jjA0376.
%
%


\begin{thebibliography}{99}{}


\bibitem{Todd-Rutel2004}
B. G. Todd-Rutel, J. Piekarewicz, and P. D. Cottle,
Phys. Rev. C \textbf{69} (2004) 021301(R) .

\bibitem{Chu2010}
Y. Chu, Z. Ren, Z. Wang, and T. Dong,
Phys. Rev. C \textbf{82} (2010) 024320.

\bibitem{Khan2008}
E. Khan, M. Grasso, J. Margueron, and Nguyen Van Giai,
Nucl. Phys. \textbf{A800} (2008) 37.

\bibitem{Grasso2009}
M. Grasso, L. Gaudefroy, E. Khan, T. Niksic,
J. Piekarewicz, O. Sorlin, Nguyen Van Giai, and
D. Vretenar,
Phys. Rev. C \textbf{79} (2009) 034318.

\bibitem{Wang11} Y. Z. Wang, J. Z. Gu, X. Z. Zhang, and J. M. Dong, Chin. Phys. Lett. 28 (2011)  102101.

 \bibitem{Nakada12} H. Nakada, K. Sugiura, and J. Margueron, arXiv:1211.5634v1 [nucl-th], 2012.

\bibitem{Burgunder2011}G. Burgunder, th\`{e}se, GANIL T 2011-06,  Universit\'{e} de Caen,
[http://tel.archives-ouvertes.fr/tel-00695010] (2011).

\bibitem{Rotaru12} F. Rotaru, F. Negoita, S. Gr{\'e}vy \emph{et al}.,
Phys. Rev. Lett. 109 (2012) 092503.


\bibitem{Yao12}
J. M. Yao, S. Baroni, M. Bender, and P.-H. Heenen,
Phys. Rev. C \textbf{86} (2012) 014310.


 \bibitem{Reinhard89}    P. G. Reinhard, Rep. Prog. Phys. \textbf{52}  (1989) 439.

 \bibitem{Ring96}        P. Ring, Prog. Part. Nucl. Phys. \textbf{37}  (1996) 193.

 \bibitem{Vretenar05}    D. Vretenar, A.~V. Afanasjev, G.~A. Lalazissis, and P. Ring,
                         Phys. Rep. \textbf{409} (2005) 101.
  \bibitem{Meng06}       J. Meng,  H. Toki, S.-G. Zhou, S. Q. Zhang, W. H. Long, and L. S. Geng, Prog. Part. Nucl. Phys. \textbf{57} (2006) 470.

\bibitem{Yao10} J. M. Yao, J. Meng, P. Ring, and D. Vretenar,
Phys. Rev. C \textbf{81} (2010) 044311.


\bibitem{Ring80} P. Ring and P. Schuck, \emph{The Nuclear Many-Body Problem}
 (Springer, Heidelberg, 1980).


\bibitem{Niksic06} T. Nik{\v{s}}i{\'c}, D. Vretenar, and P. Ring,
Phys. Rev. C \textbf{74} (2006) 064309.

\bibitem{Bender2008} M. Bender and P.-H. Heenen,
Phys. Rev. C \textbf{78} (2008) 024309.


\bibitem{Rodriguez10} T. R. Rodr{\'i}guez and J. L. Egido,
Phys. Rev. C \textbf{81} (2010) 064323.


\bibitem{Yao13} J. M. Yao, M. Bender, and P.-H. Heenen, (to be published).

\bibitem{Zhao10} P. W. Zhao, Z. P. Li, J. M. Yao, and J. Meng,
                 Phys. Rev. C \textbf{82} (2010) 054319.

 \bibitem{Krieger90} S. J. Krieger, P. Bonche, H. Flocard, P. Quentin, and M. S. Weiss,
                  Nucl. Phys. A \textbf{517} (1990) 275.

 \bibitem{Robledo09} L. M. Robledo, Phys. Rev. C \textbf{79} (2009) 021302(R).

 \bibitem{Negele70} J. W. Negele, Phys. Rev. C \textbf{1} (1970) 1260.


 \bibitem{Rychel83} D. Rychel {\em et al}., Phys. Lett. B \textbf{130} (1983) 5.

 \bibitem{Suda09} T. Suda {\em et al.,} Phys. Rev. Lett. \textbf{102} (2009) 102501.
 \bibitem{Antonov11} A. N. Antonov {\em et al.,}
                    Nucl. Instrum. Methods Phys. Res., Sect. A \textbf{637} (2011) 60.


\end{thebibliography}
\end{document}